\documentclass[review]{elsarticle}

\usepackage{lineno,hyperref}
\usepackage{color}
\modulolinenumbers[5]

\journal{Physics Letters B}

\begin{document}

\begin{frontmatter}

\title{Testing $\chi_c$ properties at BELLE II\tnoteref{mytitlenote}}
\tnotetext[mytitlenote]{Work 
supported in part by
the Polish National Science Centre, grant number DEC-2012/07/B/ST2/03867.}

\author{Henryk Czy\.z }
\author{Patrycja Kisza}
\address{Institute of Physics, University of Silesia,
PL-40007 Katowice, Poland.}

\begin{abstract}
 The model of the $\chi_{c_i}-\gamma^*-\gamma^*$\ and
  $\chi_{c_i}-J/\psi^*-\gamma^*$ form factors  developed in \cite{Czyz:2016xvc} 
  for $\chi_{c_1}$ and $\chi_{c_2}$ is extended to $\chi_{c_0}$ case.
 The studies performed within this model 
 have shown that at BELLE II it will be possible to study in detail
 $\chi_{c_i}-\gamma^*-\gamma$ form factors through measurements of
 the reaction $e^+e^-\to e^+e^- \chi_{c_i} (\to J/\psi (\to \mu^+\mu^-)\gamma)$.
 The results were obtained using the newly updated Monte Carlo generator EKHARA.
\end{abstract}

\begin{keyword}
$\chi_{c_i}$ properties, Monte Carlo generators
\end{keyword}

\end{frontmatter}

\linenumbers

\section{Introduction}

 Soon the BELLE II experiment \cite{Aushev:2010bq} will start to operate with
 unprecedented luminosity allowing to access information not available before.
 In this letter we show that 
 the integrated luminosity of 20-50 ab$^{-1}$ will
 allow BELLE II collaboration to study in detail  the $\chi_{c_i}-\gamma^*-\gamma$ form factors.
 These form factors are used in the calculations of the electronic widths
  of the $\chi_{c_i}$, which were not yet measured.  
   The theoretical
  predictions available for these widths \cite{Kuhn:1979bb,Yang:2012gk,Denig:2014fha,Kivel:2015iea,Czyz:2016xvc} depend strongly on the details of the form factors modeling and
  are different, up to two orders of magnitude, 
  even if  of all the models  agree
  with experimental data \cite{Olive:2016xmw} 
 on the $\chi_{c_i} \to J/\psi \gamma,\ i=1,2$ 
 and  $\chi_{c_2} \to \gamma \gamma$ partial decay widths.  
  We advocate here that the experimental studies of the reactions
  $e^+e^-\to e^+e^- \chi_{c_i} (\to J/\psi (\to \mu^+\mu^-)\gamma)$
  can differentiate between the proposed models.   

  To give realistic predictions for  event selections close to
   the experimental ones,
   we have extended the model developed in \cite{Czyz:2016xvc}
  to cover also $\chi_{c_0}-\gamma^*-\gamma^*$ amplitudes and 
  implemented the model amplitudes
 in the event generator EKHARA \cite{Czyz:2006dm,Czyz:2010sp}.
 The generator can also help in the data analysis of the reactions 
 $e^+e^-\to e^+e^- \chi_{c_i}$ and
 $e^+e^-\to e^+e^- \chi_{c_i} (\to J/\psi (\to \mu^+\mu^-)\gamma)$.
  The newly updated code is available from  the \href{http://prac.us.edu.pl/~ekhara/}{EKHARA} web page \\ (http://prac.us.edu.pl/$\sim$ekhara/). 
 
 The layout of this letter is the following: 
 In  Section \ref{sec1} we describe the model used in the presented simulations.
  In Section \ref{sec2} we give predictions for the expected number of events of
  the $\chi_{c_i}$ production cross sections at BELLE II and the 
 expected number of events
  for the form factor measurements. The QED non-resonant background 
 is discussed in Section \ref{sec4}. Conclusions are 
 presented in Section \ref{sec3}.

\section{The model  \label{sec1}}

 The model used in this letter is an extension of the model \cite{Czyz:2016xvc}
built to describe
 $\chi_{c_1}$ and $\chi_{c_2}$ decays to $J/\psi \gamma$, the $\chi_{c_2}$
 decay to $\gamma \gamma$ and $\psi'$ decays to $\chi_{c_{1(2)}}\gamma$. The basic assumptions used to construct the amplitudes
  for $\chi_{c_0}$ decays to $J/\psi \gamma$ and $\gamma \gamma$
 as well as $\psi'$ decay to $\chi_{c_{0}}\gamma$
 are the same as in \cite{Czyz:2016xvc}. We start from the 
$\chi_{c_i}-\gamma^*-\gamma^*$ amplitudes
 calculated in \cite{Kuhn:1979bb}  and
  assume that the Lorentz structure, as well as the form factor, are identical
  also for $\chi_{c_i}-J/\psi^*-\gamma^*$ amplitude. We allow only 
 for different coupling constants.
  From these assumptions one gets the following amplitudes for 
 the decays $\chi_{c_0}\to \gamma\gamma$, $\chi_{c_0}\to J/\psi\gamma$
 and $\psi'\to \chi_{c_0}\gamma $

   \begin{eqnarray}
&&\kern-25ptA_{0\gamma \gamma}^{\alpha\beta}(p_1,p_2)\epsilon^1_{\alpha}\epsilon^2_{\beta}\bigg{|}_{p_1^2=p_2^2=0}\kern+27pt= c_{\gamma}^0 A(p_1,p_2) , \nonumber\\
  &&\kern-25ptA_{0\gamma J/\psi}^{\alpha\beta}(p_1,p_2)\epsilon^1_{\alpha}\epsilon^2_{\beta}\bigg{|}_{p_1^2=0,\ p_2^2=M_{J/\psi}^2}\kern-10pt= c_{J/\psi}^0 A(p_1,p_2)
 ,\nonumber\\
&&\kern-25ptA_{\psi'0\gamma}^{\alpha\beta}(p_1,p_2)\epsilon^1_{\alpha}\epsilon^2_{\beta}\bigg{|}_{p_1^2=0,\ p_2^2=M_{\psi'}^2}\kern-8pt=c_{\psi'}^0 A(p_1,p_2) ,
 \label{amp0}
\end{eqnarray}
   where $\epsilon^i \equiv \epsilon(p_i)$ are the appropriate polarisation vectors,

\begin{eqnarray}
  A(p_1,p_2) &=& \frac{2}{\sqrt{6}M_{\chi_{c0}}}\left[(\epsilon^1\epsilon^2)(p_1p_2)
-(\epsilon^1p_2)(\epsilon^2p_1)\right] \left[M_{\chi_{c0}}^2 + (p_1p_2)\right],\nonumber\\
  c^0_{\gamma} &=& \frac{16 \pi \alpha}{\sqrt{m}} 
	    \cdot (a + \frac{f \cdot a^0_J}{M^2_{J/\psi}} + \frac{f' \cdot a^0_{\psi`}}{M^2_{\psi'}})
	    \cdot \frac{1}{( M^2_{\chi_{c0}}/4 + m^2)^2}\, ,\nonumber \\
   c^0_{J/\psi} &=& \frac{4 \cdot e \cdot  a^0_J}{\sqrt{m}}
	    \cdot \frac{1}{( M^2_{\chi_{c0}}/4 + m^2 -  M^2_{J/\psi}/2)^2}\, ,\nonumber \\
   c^0_{\psi'} &=& \frac{4 \cdot e \cdot  a^0_{\psi`}}{\sqrt{m}}
	    \cdot \frac{1}{( M^2_{\chi_{c0}}/4 + m^2 -   M^2_{\psi'}/2)^2}\, ,
 \nonumber \\
 \label{ff0}
\end{eqnarray}
with $f=\sqrt{\frac{3\Gamma_{J/\psi\rightarrow e^+ e^-}M_{J/\psi}^3}{4\pi\alpha^2}}$
and  $ f'=\sqrt{\frac{3\Gamma_{\psi'\rightarrow e^+ e^-}M_{\psi'}^3}{4\pi\alpha^2}}$.
 Most of the couplings are defined in \cite{Czyz:2016xvc}: 
  $a$ is proportional to the derivative of the 
 wave function at the origin, $m$ is the effective charm quark mass,
  $a_J$ and $a_{\psi`}$ are the couplings of $J/\psi-\chi_{c_i}-\gamma$ 
 and $\psi'-\chi_{c_i}-\gamma$ ($i=1,2$; not appearing in Eq.(\ref{ff0})).
$a^0_J$ and $a^0_{\psi`}$ denote the couplings of $J/\psi-\chi_{c_0}-\gamma$ 
 and $\psi'-\chi_{c_0}-\gamma$ respectively.

The coupling constants can be extracted from the experimental
data adding to the ones used in \cite{Czyz:2016xvc} the following widths

\begin{eqnarray}
\Gamma(\chi_{c0} \rightarrow \gamma\gamma) &=& \frac{3}{128\pi} |c^0_{\gamma}|^2 M^{5}_{\chi_{c0}} \, ,\nonumber \\
  \Gamma(\chi_{c0} \rightarrow J/\psi\gamma) &=& \frac{1}{192\pi} |c^0_{J/\psi}|^2 M^{5}_{\chi_{c0}}
	    (3-x)^2(1-x)^3\, ,\nonumber \\
  \Gamma(\psi' \rightarrow \chi_{c0} \gamma) &=& \frac{1}{576\pi} |c^0_{\psi'}|^2 
            (1-y)^3(1-3y)^2 \frac{M_{\psi'}^5}{y}\, ,\nonumber \\
 \label{gam0}
\end{eqnarray}
where $x =M_{J/\psi}^2/M^2_{\chi_{c_0}}$ and $y = M^2_{\chi_{c_0}}/M_{\psi'}^2$.

\begin{table}[h]
\begin{center}
\vskip0.3cm
\begin{tabular}{|c|c|c|c|c|c|}
\hline
 a[GeV$^{5/2}$]  & $m$ [GeV]& $a_J$[GeV$^{5/2}$]& $a^0_J$[GeV$^{5/2}$] & $a_{\psi'}$ [GeV$^{5/2}$]& $a^0_{\psi'}$ [GeV$^{5/2}$]\\
\hline
 0.0796   & 1.67 & 0.129 & 0.073 & -0.078 & 0.122\\
\hline
\end{tabular}
\begin{tabular}{|c|c|c|c|}
\hline
 widths [MeV] & $\chi_{c_0}$ & $\chi_{c_1}$ & $\chi_{c_2}$   \\
\hline
\hline
$\Gamma(\chi\rightarrow \gamma \gamma)_{th}$ & $2.24 \cdot10^{-3}$ &- &$5.46 \cdot10^{-4}$\\
$\Gamma(\chi\rightarrow J /\psi \gamma)_{th}$ &$1.34 \cdot 10^{-1}$ & $2.82 \cdot 10^{-1}$ & $3.74\cdot 10^{-1}$\\
$\Gamma(\psi'\rightarrow \chi \gamma)_{th}$ & $2.96 \cdot 10^{-2}$& $ 2.88 \cdot 10^{-2}$ & $2.64\cdot 10^{-2}$\\
\hline
$\Gamma(\chi\rightarrow \gamma \gamma)_{exp}$ & $2.3(2) \cdot10^{-3}$& - &$5.3(4) \cdot10^{-4}$\\
$\Gamma(\chi\rightarrow J /\psi \gamma)_{exp}$ &$1.3(1) \cdot 10^{-1}$ & $2.8(2) \cdot 10^{-1}$ & $3.7(3)\cdot 10^{-1}$\\
$\Gamma(\psi'\rightarrow \chi \gamma)_{exp}$  & $2.96(11) \cdot 10^{-2}$& $2.8(1) \cdot 10^{-2}$ & $2.7(1)\cdot 10^{-2}$\\
\hline
\end{tabular}

\caption{Model parameters and theoretical ($th$) (this paper, see also \cite{Czyz:2016xvc}), and experimental ($exp$) \cite{Olive:2016xmw} values of $\Gamma(\chi_{c_{0,1,2}}\rightarrow \gamma \gamma, \gamma J/\psi )$ and $\Gamma(\psi'\rightarrow \chi_{c_{0,1,2}}\gamma )$.}
\label{fit24}
\end{center}

\end{table}
\begin{table}[ht]
\begin{center}
\vskip0.3cm
\begin{tabular}{|c|c|c|c|c|}
  \hline
  &\cite{Czyz:2016xvc}&\cite{Kivel:2015iea}&\cite{Yang:2012gk}&\cite{Denig:2014fha}\\
  \hline
  $\Gamma(\chi_{c_1}\rightarrow e^+ e^-)$ [eV] & 0.43&0.046&0.367&0.1\\
  $\Gamma(\chi_{c_2}\rightarrow e^+ e^-)$  [eV] & 4.25&0.037&0.137& -\\
\hline
\end{tabular}
\caption{Predictions of the electronic widths of the $\chi_{c_1}$ and $\chi_{c_2}$ charmonia within recently published models. }
\label{tab11}
\end{center}
\end{table}
 The fit to 8 experimental values 
  ($\Gamma(\chi_{c_0}\rightarrow \gamma \gamma)$,
  $\Gamma(\chi_{c_2}\rightarrow \gamma \gamma)$,
  $\Gamma(\chi_{c_{i}}\rightarrow J /\psi \gamma),\ i=0,1,2$,
  $\Gamma(\psi'\rightarrow \chi_{c_{i}} \gamma),\ i=0,1,2 $ ) 
  with 6 model parameters ($a$, $m$, $a_J$, $a^0_J$, $a_{\psi'}$ and $a^0_{\psi'}$) gives $\chi^2=0.94$. The Eqs.(26-30) from \cite{Czyz:2016xvc} and 
 Eqs. \ref{gam0} from this letter were used as model predictions.
 The fit results are
  summarised in Table \ref{fit24}.

 The model parameters describing $\chi_{c1}$ and $\chi_{c2}$ are very close
  to the model parameters obtained in  \cite{Czyz:2016xvc} and the 
  predictions for the electronic widths ($\Gamma(\chi_{c_{1}} \to e^+ e^- ) = 0.37 $ eV and $\Gamma(\chi_{c_{2}} \to e^+ e^- ) = 3.86 $ eV) did change within 
  the parametric uncertainty of the model, which is about 10\%.
  The $a^0_{\psi'}$ coupling is positive at difference with the 
 negative $a_{\psi'}$.
  Both signs are the only ones allowed by the fit.

  The $\chi_{c1}$ and $\chi_{c2}$ electronic widths are calculated as loop
  integrals (see Fig.5 of \cite{Czyz:2016xvc}), thus the 
  $\chi_{ci}-\gamma^*-\gamma^*$,  $\chi_{ci}-J/\psi^*-\gamma^*$ and $\chi_{ci}-\psi'^*-\gamma^*$ form factors
  are crucial for the theoretical predictions. Table
 \ref{tab11} summarises the situation. All the models referenced there give correct predictions for
 the $\chi_{c_i} \to J/\psi \gamma,\ i=1,2$ 
 and  $\chi_{c_2} \to \gamma \gamma$ partial decay widths. Yet, the 
 predictions for the $\chi_{c1}$ and $\chi_{c2}$ electronic widths are
 different up to two orders of magnitude, showing why the experimental
  studies of the  $\chi_{ci}-\gamma^*-\gamma^*$, 
   $\chi_{ci}-J/\psi^*-\gamma^*$ and $\chi_{ci}-\psi'^*-\gamma^*$ form factors are important. 
  The $\psi'$ contribution is taken
  into account only in \cite{Kivel:2015iea}  and \cite{Czyz:2016xvc}.
  The contributions taken into account in  \cite{Kivel:2015iea}  
  and \cite{Czyz:2016xvc} are qualitatively the same, yet the differences
  coming
   from loop integrals are striking and require further studies.

\section{The amplitudes and the cross section \label{sec2}}

\begin{figure}[h]
\begin{center}
\includegraphics[width=9.cm,height=5.5cm]{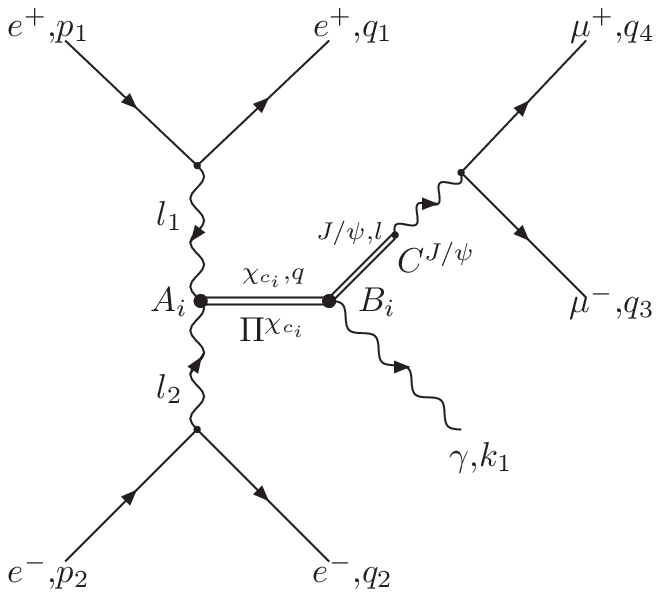}
\caption{The Feynman diagram for the amplitude of the reaction
   $e^+ e^-\to e^+ e^- J/\psi (\to \mu^+ \mu^-) \gamma$. The notation of
  four momenta is used in the formulae presented in this letter. 
\label{fd}
}
\end{center}
\end{figure}

With the couplings obtained from the fit one can predict the rates
for the reactions $e^+e^-\to e^+e^-\chi_{c_i}$ and
  $e^+e^-\to e^+e^-\chi_{c_i}(\to J/\psi(\to \mu^+\mu^-) \gamma)$.
  We do consider here only signal
 processes with the Feynman diagram given in Fig. \ref{fd}. The QED 
 non-resonant background
 can be suppressed by requiring that the $\mu^+ \mu^- \gamma$ invariant mass
 is close to the $\chi_{ci}$ mass and $\mu^+ \mu^-$ invariant mass close to 
 the $J/\psi$ mass. Its size is estimated in Section \ref{sec4}.
  As the $\chi_{ci}$
 and $J/\psi$ are almost on-shell we use a constant $\chi_{ci}-J/\psi-\gamma$
  form factor.

 The relevant amplitudes, with the four momenta denoted in Fig.\ref{fd},  read

\begin{eqnarray}
      &&\kern-30pt M_{0} =  e^3 V_{\nu}(p_1,q_1)\
    A_0^{\nu\mu}(l_1,l_2) \ \ U_{\mu}(p_2,q_2)   W^\beta(q_3,q_4) \  \Pi^{\chi_{c0}}(q) \ B_0^{\sigma}(l,k_1) \ C_{\sigma\beta}^{J/\Psi}(l^2)\nonumber \\
      &&\kern-30pt M_{1} =  e^3  V_{\nu}(p_1,q_1) \ 
   A_1^{\nu\mu\omega}(l_1,l_2) \ U_{\mu}(p_2,q_2)    W^\beta(q_3,q_4) \ \Pi^{\chi_{c1}}_{\omega\delta}(q) \ B_1^{\delta\sigma}(l,k_1) \ C_{\sigma\beta}^{J/\Psi}(l^2) \nonumber \\
      &&\kern-30pt M_{2} =  e^3 V_{\nu}(p_1,q_1) \ A_2^{\nu\mu\omega\delta}(l_1,l_2) U_{\mu}(p_2,q_2) W^\beta(q_3,q_4) \Pi^{\chi_{c2}}_{\omega\delta\pi\xi}(q)B_2^{\pi\xi\sigma}(l,k_1)C_{\sigma\beta}^{J/\Psi}(l^2) \nonumber \\     
\label{amp}
\end{eqnarray}
with  
\begin{eqnarray}
&&\kern-100ptA_0^{\nu\mu}(l_1,l_2) =   \frac{2\tilde c^0_{\gamma}((l_1-l_2)^2)}{\sqrt{6}M_{\chi_{c0}}} \left[\left(g^{\nu\mu}(l_1 \cdot l_2)-l_1^{\mu}l_2^{\nu}\right)(M^2_{\chi_{c0}}+ l_1 \cdot l_2)-g^{\nu\mu} l_1^2 l_2^2\right] , \ \nonumber \\
B_0^{\sigma}(l,k_1) &=&   \frac{2c^0_{J/\psi}}{\sqrt{6}M_{\chi_{c0}}} F^{1\sigma\nu}(k_1) l_{\nu}(M^2_{\chi_{c0}}+ k_1 \cdot l) , \ \nonumber \\
&&\kern-100ptA_1^{\nu\mu\omega}(l_1,l_2) = -i\tilde c^1_{\gamma}((l_1-l_2)^2) \left(\epsilon^{\bar{\nu}\nu\mu\omega}
\left(l_{2\bar{\nu}} l_{1}^2 -  l_{1\bar{\nu}} l_{2}^2\right) +
\epsilon^{\bar{\mu}\bar{\nu}\mu\omega} l_{1\bar{\mu}} l_{2\bar{\nu}} l_1^{\nu}
 -\epsilon^{\bar{\nu}\bar{\mu}\nu\omega} l_{1\bar{\nu}} l_{2\bar{\mu}} l_2^{\mu}\right) , \ \nonumber \\
B_1^{\delta\sigma}(l,k_1) &=& - \frac{i}{2} c^1_{J/\psi} \left(\epsilon^{\bar{\mu}\bar{\nu}\sigma\delta} F^1_{\bar{\mu}\bar{\nu}}(k_1) l^2 - \epsilon^{\bar{\mu}\bar{\nu}\bar{\alpha}\delta} F^1_{\bar{\mu}\bar{\nu}}(k_1) l_{\bar{\alpha}} l^{\sigma}\right)  \ ,\nonumber \\
&&\kern-100ptA_2^{\nu\mu\omega\delta}(l_1,l_2) = - \tilde c^2_{\gamma} ((l_1-l_2)^2)
\sqrt{2} M_{\chi_{c2}}
\left( g^{\mu\delta} l_1^{\omega} l_{2}^{\nu} - g^{\nu\mu} l_{1}^{\omega} l_{2}^{\delta}
			  - g^{\nu\omega} g^{\mu\delta} (l_1 \cdot l_2) + g^{\nu\omega} l_{1}^{\mu} l_{2}^{\delta}\right), \ \nonumber \\
B_2^{\pi\xi\sigma}(l,k_1) &=& - c^2_{J/\psi} \sqrt{2} M_{\chi_{c2}} \left(F^{1\bar{\beta}\pi}(k_1) g^{\sigma\xi} l_{\bar{\beta}} - F^{1\sigma\pi}(k_1) l^{\xi}\right)
,\nonumber \\
\tilde c_{\gamma}^0((l_1-l_2)^2) &=& \frac{16\pi\alpha}{\sqrt{ m}} (a+\frac{fa_J^0}{M^2_{J/\psi}}+\frac{f' a_{\psi'}^0}{M^2_{\psi'}}) \frac{1}{((l_1-l_2)^2/4 - m^2 +i\epsilon)^2}
                          , \nonumber \\
\tilde c_{\gamma}^{i}((l_1-l_2)^2) &=& \frac{16\pi\alpha}{\sqrt{ m}} (a+\frac{fa_J}{M^2_{J/\psi}}+\frac{f' a_{\psi'}}{M^2_{\psi'}}) \frac{1}{((l_1-l_2)^2/4 - m^2 +i\epsilon)^2}, \ i =1,2                          , \nonumber \\
c^i_{J/\psi} &=& \frac{4 \cdot e \cdot  a_J}{\sqrt{m}}
\cdot \frac{1}{( M^2_{\chi_{c_i}}/4 + m^2 -  M^2_{J/\psi}/2)^2}, \ i =1,2 ,
\nonumber \\
 && \kern-100pt C^{J/\psi}_{\sigma\beta}(l^2) = \sqrt{\frac{3\Gamma_{J/\psi \rightarrow e^+e^-}}{\alpha \sqrt{l^2}}}\frac{g_{\sigma\beta}}{l^2-M^2_{J/\psi} + i \Gamma_{J/\psi}M_{J/\psi}}
, \ \ \ \ F^1_{\mu\nu}(k_1) = (\epsilon^1_{\mu} k_{1\nu} - \epsilon^1_{\nu} k_{1\mu} ), \ \nonumber \\
&& \kern-100pt \Pi^{\chi_{c0}}(q) = \frac{1}{q^2-M^2_{\chi_{c0}} + i \Gamma_{\chi_{c0}}M_{\chi_{c0}}}, \ \ \ \ 
\Pi^{\chi_{c1}}_{\omega\delta}(q) = \frac{g_{\omega\delta}-q_{\omega}q_{\delta}/M^2_{\chi_{c1}}}{q^2-M^2_{\chi_{c1}} + i \Gamma_{\chi_{c1}}M_{\chi_{c1}}}, \
\nonumber \\
&& \kern-100pt \Pi^{\chi_{c2}}_{\omega\delta\pi\xi}(q) = \frac{\frac{1}{2}(P_{\omega\pi}P_{\delta\xi}+P_{\omega\xi}P_{\delta\pi})-\frac{1}{3}(P_{\omega\delta}P_{\pi\xi})}{q^2-M^2_{\chi_{c2}} + i \Gamma_{\chi_{c2}}M_{\chi_{c2}}}, \ \ \ \  P_{\mu\nu} = - g_{\mu\nu} + q_{\mu}q_{\nu}/M^2_{\chi_{c2}},
\nonumber \\
&& \kern-100pt V_{\nu}(p_1,q_1)=\frac{\overline{v}(p_1)\gamma_{\nu} v(q_1)}{(q_1-p_1)^2}, \ 
 U_{\mu}(p_2,q_2)=\frac{\overline{u}(q_2)\gamma_{\mu} u(p_2)}{(q_2-p_2)^2},\ W^\beta(q_3,q_4) = \overline{u}(q_3)\gamma^{\beta} v(q_4) \nonumber \\
\end{eqnarray}
where $\epsilon^1_{\mu}$ is the photon polarisation vector.
The parts of the $\chi_{c_i}-\gamma^*-\gamma^*$ vertex  vanishing
 when contracted  with
 the $e-e-\gamma^*$ vertices are not
 shown in the above formulae.  The $A_i,\ i=0,1,2$ tensors denote 
  the $\gamma^*-\gamma^*-\chi_{c_i}^*$ vertices,
  while the $B_i, \ i=0,1,2$ tensors stand for $\gamma^*-J/\psi^*-\chi_{c_i}^*$ vertices.
  Generally more tensors structures are allowed \cite{Kuhn:1979bb}, thus the results
  obtained in this letter are specific to the adopted model.
  The amplitudes were implemented into the event generator EKHARA 
  \cite{Czyz:2010sp,Czyz:2006dm}.  Two independent codes were 
   built using two different methods of spin summations to cross check the 
 implementation. 

   In principle the amplitudes should be added coherently as the final state
  is the same for all $\chi_{c_i}^*$ intermediate states. However 
  all the amplitudes drop rapidly, when the invariant mass is a bit 
 off-resonance. At about 4 decay widths off-resonance the cross
 sections drop to 1\% of the peak values. As a result the interferences
  can be safely neglected. Yet, the detector resolution effects, typically
  of order of 10-20 MeV, can result in 'moving' the events between 
  different $\chi_{c_i}$ samples. An option of simulating simultaneously
  of all $\chi_{c_i}$ production is available in the EKHARA generator
   to facilitate this simulation. The interferences between amplitudes
  are not taken into account to speed up the calculations.

  For the phase space generation of the $e^+e^-\to e^+e^- \chi_{c_i}$ reaction
  the method used in \cite{Czyz:2010sp} for the generation of the
  phase space for reactions $e^+e^-\to e^+e^-P$ ($P=\pi^0,\eta,\eta'$)
   was adopted.
   For  the simulation of the
   reaction $e^+e^-\to e^+e^- \chi_{c_i} (\to J/\psi (\to \mu^+\mu^-)\gamma)$
   the phase space generation is split into two parts:
  the first part generates the $e^+,e^-$ and the virtual $\chi_{c_i}$ four
  momenta, while the second part generates the $\mu^+,\mu^-$ and $\gamma$
  four momenta. In the first part the virtual $\chi_{c_i}$ invariant mass
  was generated using the standard change of variables to
  absorb the Breit-Wigner peak coming from  $\chi_{c_i}$ propagator, while
  the remaining variables were generated using the same method as in
   $e^+e^-\to e^+e^- \chi_{c_i}$ reaction. In the second part the 
   invariant mass $l^2$ was generated  using the standard change of variables to
  absorb the Breit-Wigner peak coming from  $J/\psi$ propagator, the photon
  angles were generated flat in the rest frame of the virtual $\chi_{c_i}$,
  while the $\mu^+(\mu^-)$ angles were generated flat in the virtual  $J/\psi$
 rest frame. In the mode, where all 3 $\chi_{c_i}$ are generated simultaneously
   the 3-channel Monte Carlo variance reduction was used.   
 The phase space generation was  cross checked with
  an independent computer code, which uses the representation
  described in \cite{Czyz:2000wh}, with flat generation of all variables.  

  The $\chi_{c_i}$ production
  cross section  in the reaction $e^+e^-\to e^+e^- \chi_{c_i}$ (the amplitudes are easy to infer from
    Eq.(\ref{amp})) integrated over the complete phase space
   with the integrated luminosity of BELLE II of
  50 ab$^{-1}$ leads to the expected number of events of about 140M ($\chi_{c_0}$),
   4.3M ($\chi_{c_1}$) and  142M ($\chi_{c_2}$). These rates will allow
   for detailed studies of many $\chi_{c_i}$ decay modes. 
    Unfortunately
    the  measurement of electronic width
   of
   $\chi_{c_1}$   through measurement of the cross section of the 
   reaction $e^+e^-\to e^+e^- \chi_{c_1}(\to{ e^+e^-})$ 
  is out of reach as the predicted
   number of events is about 2. For $\chi_{c_2}$ the situation is a bit better
    with an expected number of events equal to
    284. Further drop is however expected as the detector does not cover
     the complete  solid angle range.

\begin{figure}[h]
\begin{center}
\includegraphics[width=12.cm,height=7.5cm]{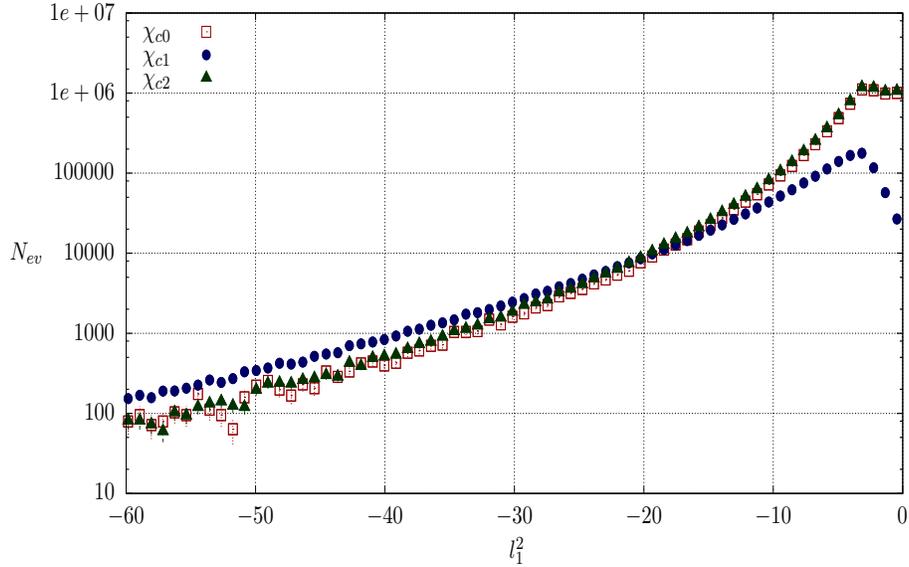}
\caption{ The distributions of expected number of events ($N_{ev}$)
   for $\chi_{c_i}$
  production, when one observes the positron in the angular range
   of 17$^\circ$ and 150$^\circ$. 
\label{prod}
}
\end{center}
\end{figure}

    If one tags a positron, in the reaction $e^+e^-\to e^+e^- \chi_{c_i}$,
   in the angular range between 17$^\circ$ and 150$^\circ$
  assuming asymmetric beams of 4 and 7 GeV with half 
  crossing angles of 41.5 mrad,
  the expected number of events drops to 6.7M($\chi_{c_0}$), 
  1.4M($\chi_{c_1}$) and 7.2M($\chi_{c_2}$).
  It shows that one has an access to
  information about  $\chi_{c_i}-\gamma^*-\gamma$ form factors. The 
   event distribution as a function virtual photon invariant mass
   ($l_1^2 = (p_1-q_1)^2$) is
   shown in Fig. \ref{prod}. When both electron and positron are observed
  in the given above angular range the expected number of events
   in the reactions $e^+e^-\to e^+e^- \chi_{c_i}$ are equal to 
   249k($\chi_{c_0}$), 174k($\chi_{c_1}$) and 295k($\chi_{c_2}$).
  It promises decent statistics in measurements 
 with doubly tagged events.
  .

 In this letter we concentrate on the possible tests of the 
  models of $\chi_{c_i}-\gamma^*-\gamma^*$ form factors using single and
  double tag events
  relying on an identification of a simple $\chi_{c_i}$ decay mode:
  $\chi_{c_i}\to J/\psi (\to \mu^+\mu^-)\gamma$.
  If one requires identification
  of $\chi_{c_i}$ and $J/\psi$ through invariant masses of $\mu^+\mu^-\gamma$
  and $\mu^+\mu^-$ final states respectively,
  the $\chi_{c_i}-J/\psi^*-\gamma$ form factors are entering the cross section
  with fixed invariants, thus they are almost constant. This way the 
  $\chi_{c_i}-\gamma^*-\gamma$ form factors can be measured.
   
  In the results presented below we assume 
 the asymmetric beams of 4 and 7 GeV with half crossing angles of 41.5 mrad.
  We assume also that the particles 
  ($\mu^+,\mu^-$, photon and positron and/or electron)
  can be detected and their four momenta measured 
 if their polar angles are between 17$^\circ$ and 150$^\circ$
   \cite{Abe:2010gxa}. 
  
   \begin{figure}[h]
\begin{center}
\includegraphics[width=12.cm,height=7.5cm]{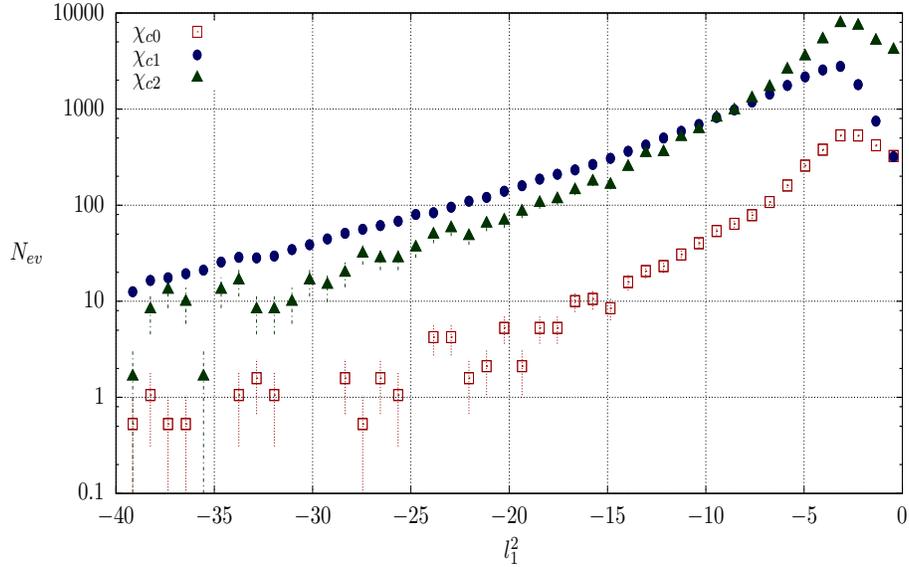}
\caption{ The distributions of expected number of events ($N_{ev}$)
   for $\chi_{c_i}$
  production with subsequent decay to $J/\psi (\to \mu^+\mu^-)-\gamma$.
 The event selection is described in the text.
\label{decay}
}
\end{center}
\end{figure}

\begin{figure}[h]
\begin{center}
\includegraphics[width=6.cm,height=6.cm]{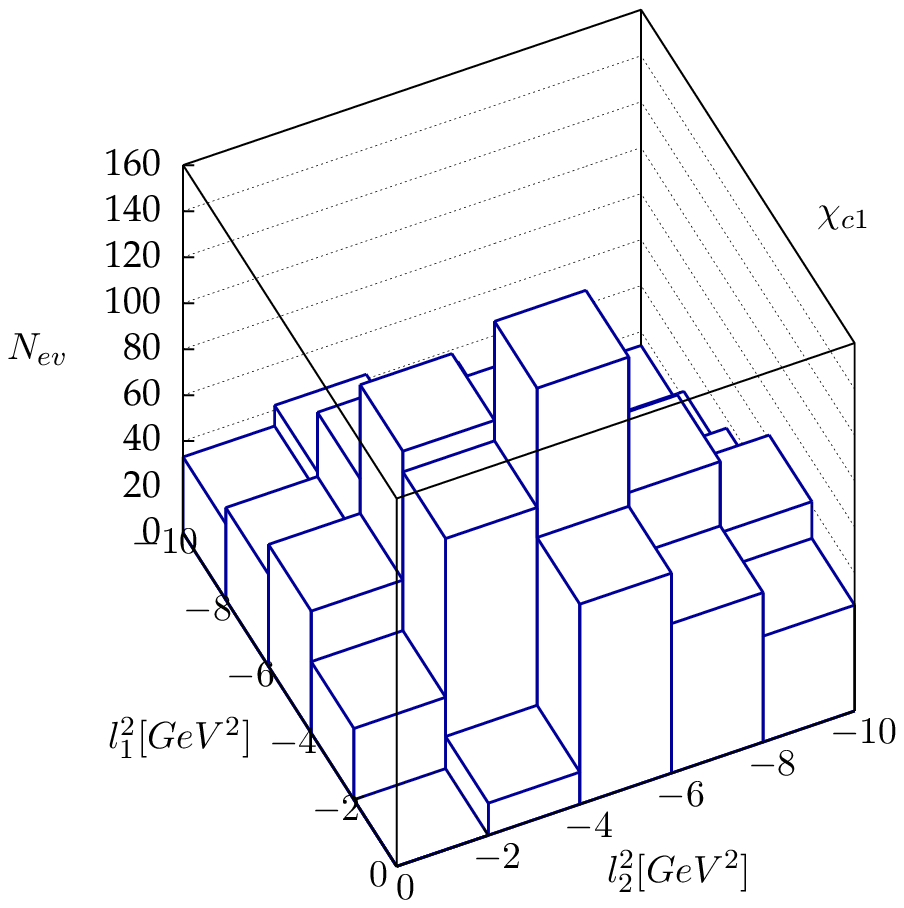}
\includegraphics[width=6.cm,height=6.cm]{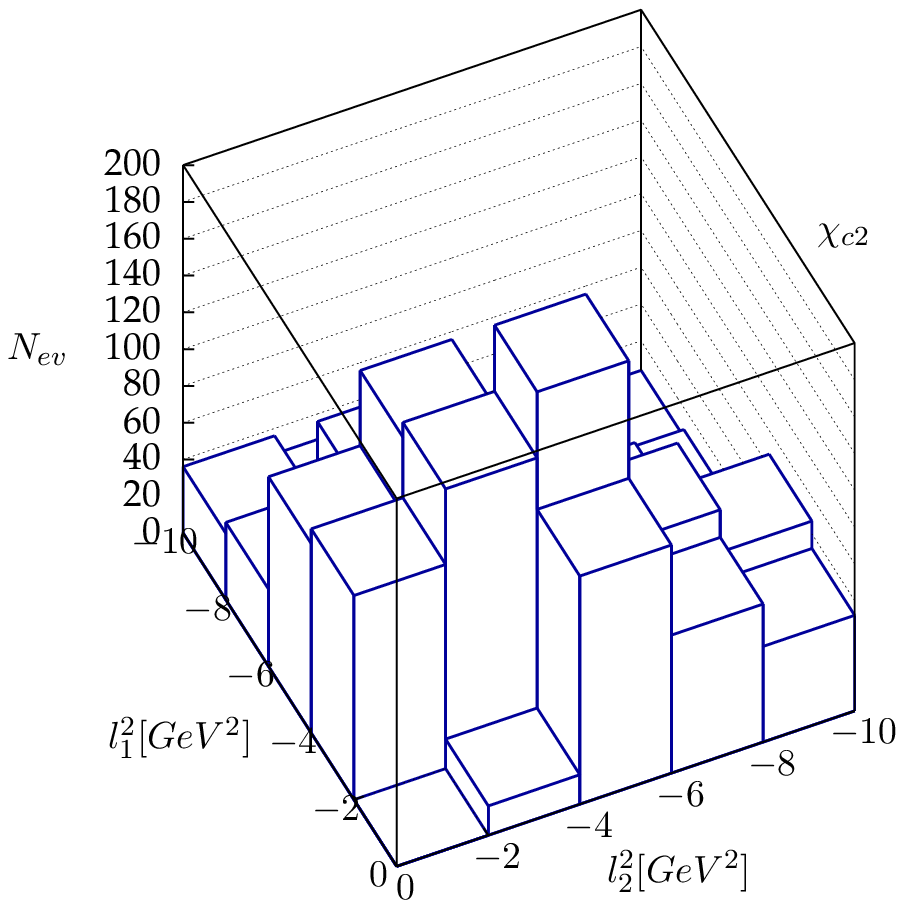}
\caption{ The distributions of expected number of events ($N_{ev}$)
   for $\chi_{c_1}$ and for $\chi_{c_2}$
   production with subsequent decay to $J/\psi (\to \mu^+\mu^-)-\gamma$
    when both electron and positron are tagged.
 The event selection is described in the text.
\label{decay-dt}
}
\end{center}
\end{figure}

 If the electron is not tagged, 
  the expected numbers of events after the applied cuts are
   3114 for $\chi_{c_0}$, 21819 for $\chi_{c_1}$ and 44126 for $\chi_{c_2}$.
   It will allow for testing of the $\chi_{c_i}-\gamma^*-\gamma$ form factors
   for the first time. The $l_1^2$ invariant mass distribution is shown
  in Fig. \ref{decay}. The $l_2^2$ invariant mass is, as expected,
  limited to small values with 2770 ($\chi_{c_0}$), 17892 ($\chi_{c_1}$)
   and 38863 ($\chi_{c_2}$) events with $-1$ GeV$^2<l_2^2 < 0$.
   Thus the form factor can be 
   extracted with a decent accuracy for one of the invariants close to zero
   and the second
  spanning up to about -30 GeV$^2$. With a limited statistics one can
  even have data for $\chi_{c_1}$ ( 2538 events) and $\chi_{c_2}$ ( 2472 events)
  with all the particles observed in the detector allowing for an accurate
    reconstruction of both invariants. The expected
     event distributions are shown in Fig.\ref{decay-dt}.
   For $\chi_{c_0}$ the expectation is 136 events.
 \vskip 1cm
     
\section{QED background estimates \label{sec4}}

The non-resonant QED background estimation was performed
  using the HELAC-PHEGAS generator \cite{Cafarella:2007pc}.
 The event selections were identical to the ones used to obtain
 the signal events.
 The ranges of $\mu^+\mu^-$ and $\mu^+\mu^-\gamma$ invariant masses were chosen
 to contain 99\% of the signal cross section: $3.0965 \le l^2 \le 3.0973$,
 $3.37 \le q^2 \le 3.50$ for $\chi_{c0}$,
 $3.50191 \le q^2 \le 3.51941$ for $\chi_{c1}$,
 $3.5475 \le q^2 \le 3.5650$ for $\chi_{c2}$.
 The polar angles of the observed particles ($\mu^+\mu^-\gamma e^+$ for single tag events and
 $\mu^+\mu^-\gamma e^+e^-$ for double tag events) were required
 to be between 17$^\circ$ and 150$^\circ$ in the laboratory frame (see Section \ref{sec2}).
  For $\chi_{c_0}$ production and decay
  the background is not negligible: 110\% for single tag events
  and 220\% for double tag events. It shows that in this case
  the interference effects between background and signal are important
  and will have to be studied. Yet, as the signal is small as compared
  to $\chi_{c_1}$ and $\chi_{c_2}$ the expected statistical accuracy
  will also be much worse. 
  For $\chi_{c_1}$ and $\chi_{c_2}$ the background to signal ratio
  is much smaller
   for two reasons: the signals are bigger by one order od magnitude
   and the decay widths are smaller about one order of magnitude
   as compared to $\chi_{c_0}$ decay width. For single tag
   events the background to signal ratio is
    0.2\% for $\chi_{c_1}$ and 0.7\% for $\chi_{c_2}$,
   while for double tag events it is 0.1\% and 1.7\% respectively.
   Thus for  $\chi_{c_1}$ and $\chi_{c_2}$ the interferences between
  background and signal amplitudes can be neglected and the background
   can be simulated using the existing Monte Carlo generators.
   The resonant background, mainly the 
   $e^+e^-\to e^+e^- J/\psi(\to \mu^+\mu^-)\gamma$ process without $\chi_{c_i}$
  intermediate states involved, should also be studied. None of the existing
  generators is currently able to generate this process and the main difficulty
  will be to find an efficient generation algorithm.

\section{Conclusions \label{sec3}}
 The model of the $\chi_{c_i}-\gamma^*-\gamma^*$,
  $\chi_{c_i}-J/\psi^*-\gamma^*$ form factors developed in \cite{Czyz:2016xvc}
  for $\chi_{c_1}$ and $\chi_{c_2}$ is extended to $\chi_{c_0}$ case.
 Within this model, it was shown that at BELLE II it will 
 be possible to study in detail
 $\chi_{c_i}-\gamma^*-\gamma^*$ form factors through measurements of
the reaction $e^+e^-\to e^+e^- \chi_{c_i} (\to J/\psi (\to \mu^+\mu^-)\gamma)$.
 It is achieved by event selections, which force the $\chi_{ci}$ and $J/\psi$ to be almost
 on-shell. For these kinematic configurations the QED background is negligible for 
 $\chi_{c1}$ and $\chi_{c2}$, while for $\chi_{c0}$ it has to be taken into account.
 The proposed measurements should clarify, which of the models giving predictions for the $\chi_{c_1}$
 and $\chi_{c_2}$ electronic widths is correct, even without direct measurement
 of these widths. If the electronic widths are measured as well, they will
 allow for further refinements of the models. 
 The expected number of events
 for the $\chi_{c_i}$ production show that detailed studies of the
  $\chi_{c_i}$ branching ratios will also be possible at BELLE II. 
 The newly updated Monte Carlo generator EKHARA can be of help
   for the visibility studies and the data analyses.

\section*{References}

\bibliography{mybibfile}

\end{document}